\documentstyle[11pt]{article}

\begin{document}
\begin{titlepage}
\begin{center}
February, 1997      \hfill     BUHEP-97-4\\
\vskip 0.2 in
{\large \bf HADRONIC STRING FROM CONFINEMENT}

\vskip .2 in
         {\bf   Raman Sundrum\footnote{email: sundrum@budoe.bu.edu.}}
        \vskip 0.3 cm
       {\it Department of Physics \\
Boston University \\
Boston, MA 02215, USA}
 \vskip 0.7 cm 

\begin{abstract}
The $existence$ of hadronic string is derived as a consequence of
confinement in QCD. A state of ``stretched
glue'', created by a Polyakov loop operator,  is shown to have translational
zero-modes which are stringy degrees of freedom. These modes are
described by an effective string theory, valid for worldsheets which are
locally flat on length  scales of order $\Lambda_{QCD}^{-1}$, the
dominant behavior being given by the Nambu-Goto
action. In a subsequent paper,
these effective strings will be shown to emerge in  mesons
stretched by their orbital angular momentum, thereby deriving some
aspects of Regge phenomenology directly from QCD and confinement.
\end{abstract}
\end{center}

\end{titlepage}

\section{Introduction}

Historically, the string model of hadrons emerged as a beautiful
realization of dual models and Regge phenomenology. Prior to the
ascendence of QCD, it was considered a prime candidate for the theory of
the strong interactions. However, as is well known, when the Nambu-Goto string
is taken as a fundamental principle, it leads to twenty-six dimensions,
gravitons, tachyons, and other disturbing consequences \cite{gsw}.
Furthermore, it fails to reproduce the partonic behavior found in
deep-inelastic scattering experiments.  

Of course, the modern viewpoint is that QCD is the fundamental theory of
the strong interactions. Strings may then provide an {\rm effective}
theory of some sector of QCD. Just like the chiral lagrangian
description of pions, the effective theory will have a restricted domain
of validity, beyond which it breaks down. A common picture is that QCD
strings are vibrations of color-electric flux tubes which connect and
confine quarks inside hadrons. Such flux tubes do indeed
appear in the strong-coupling limit of lattice QCD and are responsible
for the area law of confinement \cite{wilson}.\footnote{The
  strong-couping limit is a rather drastic approximation, but is usually
  assumed to give a reasonable {\it qualitative} guide to the physics of
  confinement.}
In the continuum, we expect that the flux-tubes will look stringy when
they are much longer than their characteristic thickness,
$\Lambda_{QCD}^{-1}$. This suggests that the effective theory is valid
for strings much longer than $\Lambda_{QCD}^{-1}$. See ref. \cite{effst}
for a discussion of effective string theory from this perspective. 
 This interpretation bypasses the
various embarrassments of fundamental hadronic strings. As a simple
example, the massless graviton of fundamental string theory is a short
closed string, and therefore is not predicted within the effective
theory.

In this paper, I show from first principles (Yang-Mills
theory) that effective string degrees of freedom must exist
 as a necessary consequence of confinement, and that the string
 description is dominated by the Nambu-Goto action.  The derivation
 clarifies and justifies the string picture within QCD.  
To illustrate the relationship between strings and confinement,
consider the analogous relationship between Coulomb's law and photons in
quantum electrodynamics. Suppose experimentally, we had only observed
the static force between electric charges, given by Coulomb's
law. However let us say that one firmly believes in the principles of
special relativity, charge conservation, and quantum mechanics. To
describe electrostatics, one could invent the concept of an electric
field satisfying Gauss' law, 
\begin{equation}
\partial_i E_i(x) = \rho(x).
\end{equation}
Now, since charge is conserved, charge-density, $\rho(x)$, must
transform as the time-component of a $4$-vector, $J_{\mu}$. Therefore to
accomodate Gauss' law in a relativistic context we must take $\vec{E}$
to be  part of an antisymmetric tensor satisfying Maxwell's equation,
\begin{equation}
\partial_{\mu} F^{\mu \nu} = J^{\nu}. 
\end{equation}
In particular, magnetic fields are necessary. Thus the static Coulomb law
leads to a new $dynamical$ object: the electromagnetic wave. In the
quantum theory the associated particle is the photon. Similarly,
reconciling Newton's law of gravitation with relativity led Einstein to
General Relativity as the unique low-energy theory of gravity. The
resulting gravitational waves appear in the (effective) quantum theory
as gravitons. To summarize, a
 static long-range force between charges always has associated waves in
 a relativistic theory.
In QCD the static force is confining. That is, it costs an infinite
energy to pull apart the quarks of a single hadron. In the real world
there is the complication that a color-singlet hadron can break into two
color-singlets given only a finite amount of energy by virtue of quark
pair production. We can postpone this problem by either considering a
world  where there are no light dynamical quarks, or working in the
large-$N_c$ limit, where pair production is suppressed. In this paper we
will do the former. As will be seen in section 2, general principles
and confinement imply a linear potential at long distances. This
restriction is
analogous to electrodynamics, where a $1/r^3$ potential, say, would be
inconsistent with general principles. What are the waves that correspond
to the confining force in QCD? I will show that they are string vibrations.

Waves on a string naturally have an interpretation as  
Nambu-Goldstone (NG) modes of the spontaneous breaking of
translational symmetry transverse to the classical string ground
state.\footnote{Quantum-mechanically this is a little trickier, because
  strictly speaking, fluctuations in this effectively
  $(1+1)$-dimensional system wipe out the symmetry breaking
  \cite{coleman} \cite{luscher}. This subtlety will be properly treated
  in the later sections.}
This strongly suggests that if we wish to {\rm derive} strings as a general
consequence of confinement, we should demonstrate that the state of a
stretched  hadron spontaneously breaks spacetime 
symmetries in the pattern for which strings are the requisite NG
modes. This is the approach taken in this paper. The hadron will be a
pure glue state created by a large Wilson loop (a ``Polyakov loop'' to
be precise,  as explained later). The symmetry-breaking will be derived
as a consequence of confinement. An important tool for describing the
resulting NG modes will be the chiral
lagrangian approach. The chiral lagrangian construction for internal
symmetry breaking by Callan, Coleman, Wess and Zumino \cite{ccwz} has
been generalized by Volkov \cite{volkov}
 to incorporate spontaneous breaking of spacetime
symmetries.  I will use a version of this
approach, streamlined for the application to strings. As in the case of
internal symmetries, the chiral lagrangian is expected to
(model-independently) enforce the constraints of unitarity and 
(spontaneously broken) symmetry on the properties of the NG modes. In
the present case the Poincare symmetries themselves are realized
non-linearly in the chiral lagrangian.

In the interests of simplicity,  the  arguments presented in
this paper  are made for pure Yang-Mills dynamics, without light
dynamical quarks.
In a second paper \cite{2nd}, I will make contact with
phenomenology, showing that strings emerge when a meson is
stretched by  orbitally exciting it. In certain kinematic regimes the
resulting Regge trajectories are linear. Though the derivation of Regge 
trajectories from
the string model is hardly new, what is new is their derivation 
directly from QCD and confinement, without model assumptions, and with an
understanding of the domain of validity. 
The large-$N_c$ approximation is
used in the second paper to focus on confinement and to treat the
breaking of strings as a finite-$N_c$ correction.

Finally, though the subject of this paper is the effective string, I
wish to  briefly mention the ongoing endeavor to demonstrate an {\it
  exact} duality between QCD and {\rm some} fundamental string
theory (differing from Nambu-Goto strings).
 The basis for this duality is the Gauss law constraint on
physical states of a gauge theory (which is also implicit in the present
derivation of the effective string). The constraint is satisfied by
eigenstates of the color-electric field where the field lines are closed
or end on quarks. Gauge theory written in terms of the evolution of
these flux-line states has the form of a string field theory. A simple
way to see this concretely  is provided by the hamiltonian lattice
formulation of QCD. See for example refs. \cite{susskind}
\cite{bander}. Unfortunately, the explicit form of the
fundamental dual string theory in the continuum is still unknown.
Ref. \cite{polchinski} provides a brief and useful review of the progress
and problems in this area, as well as references. More recently,
it has been shown that 2-dimensional large-$N_c$ QCD 
 can be exactly reformulated as a string theory \cite{2d}.

 This paper is organized as follows. Section 2 is a derivation of the
 area law (the euclidean spacetime version of a linear potential) from
 the assumption of confinement of static charges. In section 3, I
 consider  glue ``wrapped around the universe'' and show that there
 is ``almost'' a spontaneous symmetry-breaking of transverse
 translations as a direct consequence of the area law. This is
 sufficient  to derive  an analog of  Goldstone's theorem,
 demonstrating the existence
 of stringy NG modes. This is done in section 4. In section 5, I
 construct the chiral lagrangian description of the NG modes, dominated
 by the Nambu-Goto term.  Section 6 provides discussion as well as
 connections to the second paper and future work.

\section{The Area Law}

For the remainder of the paper we consider the euclidean spacetime
formulation of Yang-Mills theory for $SU(N_c)$ gauge group, with $N_c$
finite (say 3). 
In this section the area law is derived as the only possible static law of
confinement. It is the euclidean reflection of a linear
potential. Confinement is approached in the standard way, by
considering the response to  test charges.
 Consider a meson made from a
static (infinitely massive) quark and anti-quark pair, 
both at rest in some frame.
The dynamics is described by a heavy quark effective lagrangian,
\begin{equation}
{\cal L} = \bar{Q}_+  D^+_t Q_+ ~+~ \bar{Q}_-  D^-_t Q_-
 + \frac{1}{4 g^2} {\rm tr} G_{\mu \nu} G_{\mu
  \nu},
\end{equation}
where $Q_{+/-}$ are the quark and anti-quark fields.
In this static limit the quarks can have fixed positions, with
separation $L$. We are using the static quarks to stretch a meson and
examine confinement, so  consider $L \gg \Lambda_{QCD}^{-1}$.
The dynamical object is the state of glue connecting the
quarks. The energy eigenvalue of the lowest such state, $E_0(L)$, is
then naturally identified as the  potential energy needed to
achieve the quark separation, $L$. Confinement
corresponds to the statement,
\begin{equation}
E_0(L)~~ \rightarrow_{L \rightarrow \infty} ~~\infty.
\end{equation}

 A gauge-invariant interpolating operator for the meson (which is
necessarily non-local due to the fixed quark separation) is given by
\begin{equation}
{\cal O}(t) \equiv Q_+(0, t) P e^{i \int_0^L dx A_x(x,t)} Q_-(L,t),
\end{equation}
where $y$  and $z$ are fixed on the right-hand side, so I have
suppressed writing them. Remembering the rule for going from Minkowski
to euclidean time-evolution, $e^{iHt} \rightarrow e^{-Ht}$, the spectral
decomposition of the ${\cal O}$ correlator for $t > 0$ reads,
\begin{equation}
<{\cal O}(t) {\cal O}^{\dag}(0)> ~=~ \int_{E_0(L)}^{\infty} dE~ \rho(E, L)~ 
e^{-Et}, 
\end{equation}
where $\rho$ is the spectral density,
\begin{equation}
\rho(E, L) ~\equiv~ \sum_r |<0|{\cal O}(0)|r,L>|^2 ~\delta(E - E_r),
\end{equation}
and the $|r,L>$ constitute the complete energy basis for the
glue in the meson. For very large euclidean times the correlator is
 dominated by the lowest-lying intermediate state of glue.

 The position-space static quark propagator in the gauge field
background is given by the  identity,
\begin{equation}
 D_t~ [\theta(t) \delta^3(\vec{x}) P e^{i \int_0^t dt'
  A_t(\vec{x}_0, t')}] ~=~  \delta^4(x).
\end{equation}
Using it, we can integrate out the quarks (static
  quarks having unit functional determinant) in  the ${\cal  O}$
 correlator, giving
\begin{equation}
<{\cal O}(t) {\cal O}^{\dag}(0)> ~~=~~ <{\rm tr} P e^{i\int_C
  dx_{\mu} A_{\mu}}>,
\end{equation}
where $C$ is a rectangular loop with sides of length $t$ and $L$. 
 In deriving this equation 
a trivial factor of  $(\delta^3(0))^2$ has been dropped by a renormalization 
of ${\cal O}$, making ${\cal O}$ dimensionless. Combining this
result with the spectral decomposition, we have,
\begin{equation}
 <{\rm tr} P e^{i\int_C  dx_{\mu} A_{\mu}}>~~~ \sim_{t \rightarrow \infty}
 ~~ e^{-E_0(L) t}.
\end{equation}
Now the left-hand side has an obvious symmetry to exploit
 to get information about $E_0(L)$, namely
it is invariant under a $90$-degree rotation of the rectangle, $C$,  combined
with exchanging $t$ and $L$. This is just a consequence of
$4$-dimensional euclidean rotational symmetry. One complication in using
this is that,  since there
are two length scales, $L \gg \Lambda_{QCD}^{-1}$,
 it is not immediately obvious
whether the above asymptotic  behavior in $t$ sets in for $t \gg
\Lambda_{QCD}^{-1}$, or much later,  say for  $t \gg L$. I will show that
it is  the former case, if I make the  extremely plausible technical
assumption that ${\cal O}$ is a ``good'' interpolating operator, as
defined below. 

First note that in the $t \rightarrow 0$ limit of eqs. (6) and (9) we find,
\begin{eqnarray}
 \int_{E_0(L)}^{\infty} dE~ \rho(E, L) ~&=& ~< {\rm tr} 1>~ \nonumber \\
&=&~ N_c.
\end{eqnarray}
The precise technical assumption I need to make is that there is some
non-zero fraction $k$, such that for sufficiently large $L$,
\begin{equation}
 \int_{E_0(L)}^{E_0(L) + \Lambda_{QCD}} dE~ \rho(E, L) ~>~ k N_c.
\end{equation}
In words, ${\cal O}^{\dag}|0>$ has a non-vanishing overlap with 
low-lying states of glue for large $L$.  Given
this assumption, two simple bounds follow,
\begin{eqnarray}
 \int_{E_0(L)}^{E_0(L) + 2 \Lambda_{QCD}} dE ~\rho(E, L)~ e^{-Et} ~&>&~
 \int_{E_0(L)}^{E_0(L) + \Lambda_{QCD}} dE ~\rho(E, L) ~e^{-Et}
 \nonumber \\ 
~&>&~ k N_c 
~e^{- [E_0(L) + \Lambda_{QCD}] t}, \nonumber \\
 \int_{E_0(L) + 2 \Lambda_{QCD}}^{\infty} dE ~\rho(E, L)~ e^{-Et} ~&<&~ (1
 - k) N_c~ e^{- [E_0(L) + 2 \Lambda_{QCD}] t}.
\end{eqnarray}
For $t \gg \Lambda_{QCD}^{-1}$,  we see that the  states between
$E_0$ and $E_0 + 2 \Lambda_{QCD}$  dominate. Therefore, 
\begin{equation}
 <{\rm tr} P e^{i\int_C  dx_{\mu} A_{\mu}}>~ ~\sim_{t \gg 
\Lambda_{QCD}^{-1}} ~ 
~ \int_{E_0(L)}^{E_0(L) + 2 \Lambda_{QCD}} dE ~\rho(E, L)~ e^{-Et}.
\end{equation}
Since confinement implies $E_0 \gg 2 \Lambda_{QCD}$ for large $L$, we
finally arrive at 
\begin{equation}
 <{\rm tr} P e^{i\int_C  dx_{\mu} A_{\mu}}>~~ ~\sim_{t, L \gg
   \Lambda_{QCD}^{-1}} ~ ~  e^{-E_0(L) t}.
\end{equation}

We can now straightforwardly use euclidean rotational symmetry to  exchange 
``time'' and ``length'' for the meson. That is, in euclidean space we can
equally well choose the direction parallel to the $L$-length side of the
rectangle to be ``time'' and interpret $t$ to be the inter-quark separation. 
Repeating all our steps we find,
\begin{equation}
 <{\rm tr} P e^{i\int_C  dx_{\mu} A_{\mu}}> ~\sim_{t, L \gg
   \Lambda_{QCD}^{-1}} ~  e^{-E_0(t) L}.
\end{equation}
Comparing eqs. (15) and (16), 
\begin{equation}
 e^{-E_0(L) t} ~\sim_{t, L \gg   \Lambda_{QCD}^{-1}} ~  e^{-E_0(t) L}.
\end{equation}
This has a unique confining asymptotic solution, 
\begin{equation}
E_0(L) ~~\sim_{L \gg \Lambda_{QCD}^{-1}} ~~~\Lambda_{QCD}^2 L,
\end{equation}
which completes the proof that the static potential is linear at long 
distances.\footnote{I have used ``$\Lambda_{QCD}$'' a few times to
  denote different quantities which are all parametrically set by the
  strong interaction scale. The different ${\cal O}(1)$ prefactors are
  unimportant.} Eq. (15) then 
tells us that the expectation of a large rectangular
Wilson loop is exponentially suppressed by its area. This is the
standard statement of the area law of confinement. 

\section{``Symmetry-Breaking'' by Stretched Glue}

In the last section we obtained a stretched state of glue by pinning the
ends down with infinitely massive quarks. These special endpoints are a
nuisance, so in this section we will do away with them. (Of course
phenomenologically dynamical quarks are attached at these ends, but this
complication is deferred until the second paper.)  Glue does not
feel any other force, so to stretch it we will use spacetime itself. To
this end, consider the spatial $x$-direction to be compact, with a large
radius, $R \gg \Lambda_{QCD}^{-1}$. Imagine taking the static quark ends of
the meson of the last section to be further and further apart in the
$x$-direction. As this is done, the energy of the glue in between grows
linearly, as we saw. Finally, the static quarks come together after
winding ``around the universe'', since the
$x$-direction is compact, with the lowest glue energy-eigenvalue being $2
\pi R \Lambda_{QCD}^2$, presumably. Let us now annihilate the quarks,
leaving a state of pure glue wrapped around the $x$-axis. The
glue cannot contract without breaking (and there are no light quarks so
this is disallowed), so it remains conveniently 
stretched around the $x$-direction. Presumably its energy remains
$2 \pi R \Lambda_{QCD}^2$.  However now that it is free of the static quarks,
 it is  completely free to move in the  $y$
and $z$ directions. From the
viewpoint of the effective $(2+1)$-dimensional theory below the
compactification scale, $1/R$, the winding glue state looks like a point-like
glueball, which is very heavy compared to $\Lambda_{QCD}$. 

To precisely state and prove the above intuitions, we proceed as
follows. An interpolating operator for the winding glueball state is
provided by a Polyakov loop, which is a winding Wilson loop operator
parallel to the $x$-axis. It is given by 
\begin{equation} 
{\cal P}(y, z, t) ~\equiv~~ {\rm tr} P e^{i \int_0^{2 \pi R} dx A_x(x,
  y, z, t)}. 
\end{equation}
The intermediate states in the spectral decomposition of the ${\cal
  P}$-correlator are the winding glue states of interest: 
\begin{eqnarray}
<{\cal P}(y,z,t) {\cal P}^{\dag}(0,0,0)>~ &=&
\int_{s_0(R)}^{\infty} 
ds~ \int \int  \frac{dp_y dp_z}{2 \sqrt{s + p^2_y + p^2_z}}
 \rho(s, R) \nonumber \\
&~& \times e^{i(p_y y + p_z z)} ~e^{-\sqrt{s + p_y^2 + p_z^2} t}. 
\end{eqnarray}
The spectral density is a function of the $(2+1)$-dimensional 
mass-squared of the
intermediate states, $s$, since I have used translation symmetry in the
$y$ and $z$ directions to sum the corresponding momentum eigenstates
explicitly,  
\begin{equation}
\rho(s, R) ~\equiv~ \sum_a 
|<0|{\cal P}^{\dag}(0,0,0)|a,R>|^2 ~\delta(s - m^2_a).
\end{equation}
The fact that I have used $(2+1)$-dimensional notation is a matter of
convenience. The formulas are exact and not low-energy approximations. 
The fact that the world is fundamentally 
$(3+1)$-dimensional should be reflected in
the spectrum of glue states, which is as yet undetermined. 
If our earlier
 intuition is correct the correlator should obey the area law of
confinement since all the intermediate states are stretched glueballs,
\begin{equation}
<{\cal P}(y,z,t) {\cal P}^{\dag}(0,0,0)> ~~\sim_{R, \sqrt{t^2 +
    y^2 + z^2} \gg \Lambda_{QCD}^{-1}} ~~ e^{- 2 \pi R
  \sqrt{t^2 + y^2 + z^2} \Lambda_{QCD}^2}.
\end{equation}
The proof follows by using the old trick of exchanging euclidean 
time with the $x$-direction. That is, we
can interpret the compact direction as an imaginary ``time'' and take
all the non-compact directions to be spatial. This puts us in the
framework of finite-temperature field theory, with $\beta \equiv 2 \pi
R$. The ${\cal P}$-correlator is now precisely the finite temperature partition
functional for the mesonic glue states of the
previous section, where the static quarks are separated by 
\begin{equation}
L \equiv \sqrt{t^2 + y^2 + z^2}.
\end{equation}
 By turning our heads by $90$ degrees, the
zero-temperature problem of winding glue has transformed into a
finite-temperature problem for the static-quark mesons! The two Polyakov
loops are now interpreted as the world-lines for the static quarks of
the meson, which are of course color sources for glue. Thus, 
\begin{equation} 
<{\cal P}(y,z,t) {\cal P}^{\dag}(0,0,0)> ~~ = ~\sum_r e^{- \beta
  E_r(L)}. 
\end{equation}
(I am distinguishing the states of glue stretched between static quarks
 from the
states of glue wrapped around a compact spatial direction, by denoting them
$|r>$ and $|a>$ respectively.) For temperatures much smaller than 
$\Lambda_{QCD}$, the system is dominated by the lowest-lying
states. We
found in the last section that their energies scale like $\Lambda_{QCD}^2
L$ for $L \gg \Lambda_{QCD}^{-1}$. Therefore, writing $\beta$ and $L$
in terms of the original variables, $R, y, z, t$, we arrive at the area
law for Polyakov loops, eq. (22).

We can deduce the nature of the lowest-lying states $|a, R>$, 
interpolated by ${\cal
  P}$, by Fourier transforming the area law with respect to $y$ and
$z$ (but not $t$).
 Since small and large $y$ and $z$ must be considered, we must
take $t \gg \Lambda_{QCD}^{-1}$ in order to remain in the domain of
validity of the area law.   
We get, 
\begin{eqnarray}
<{\cal P}(t) {\cal P}^{\dag}(0)>(p_y, p_z) &\sim_{R, t \gg
  \Lambda_{QCD}^{-1}}& \int dy dz~ e^{ip_yy + ip_zz} e^{-m_0 \sqrt{t^2 +
    y^2 + z^2}} \nonumber \\
&\sim& \int dy dz~  e^{ip_yy + ip_zz} e^{-m_0 t(1 + \frac{y^2 + z^2}{2 t^2})}
\nonumber \\
&\sim&  e^{-(m_0 + \frac{p_y^2 + p_z^2}{2 m_0})t},
\end{eqnarray}
where,
\begin{equation}
m_0 \equiv 2 \pi R \Lambda_{QCD}^2. 
\end{equation} 
Comparing this result with the spectral decomposition, we
deduce that for $R \gg \Lambda_{QCD}^{-1}$  the lowest-lying states created
by the Polyakov loop  are the $|m_0; p_y, p_z>$, with mass eigenvalue
$m_0$. The derivation has implicitly expanded for $p_y, p_z \ll m_0$,
but by $(2+1)$-dimensional relativistic invariance, the exact dispersion
relation is given by,
\begin{equation}
E_0 = \sqrt{m_0^2 + p_y^2 + p_z^2}.
\end{equation}

The  existence of the $|m_0; p_y,
p_z>$ states is the phenomenon of ``almost''  spontaneous symmetry-breaking of
transverse translations
referred to in the introduction, in the sense that states with  fixed
momenta $(p_y, p_z)$, become degenerate with the winding ground state 
$|m_0, p_y = p_z = 0>$, in the $R \rightarrow \infty$ limit. Strictly
there is no symmetry-breaking\footnote{This is true even in the $R
  \rightarrow \infty$ limit \cite{luscher}.} 
since $|m_0, p_y = p_z = 0>$ is
translation invariant, but the near degeneracy will suffice to
demonstrate the presence of gapless ``Nambu-Goldstone'' modes with non-zero
$x$-momentum, by means
closely analogous to the proof of Goldstone's theorem. This is done in
the next section. 

Before going to the formal derivation it is useful to build some
intuition. 
We can summarize our result thus far by writing an effective theory for the
winding glue states we have found,
\begin{equation}
S = 2 \pi R \Lambda_{QCD}^2 \int dt  \sqrt{1 + (\partial_t
y)^2 + (\partial_t z)^2}. 
\end{equation}
 For momenta below the compactification scale $1/R$,
this effective theory looks very reasonable. The world is effectively
$(2+1)$-dimensional and we see a glueball state described by the
action for a relativistic point-particle of mass $m_0 = 2 \pi R
\Lambda_{QCD}^2$. There is nothing strange about the presence of the
ultraviolet scale $R$ in the lagrangian. However, our result is also
valid for $p_y, p_z \gg 1/R$, and therefore must somehow be part of a
fully $(3+1)$-dimensional theory. From this perspective eq. (28) seems
more disturbing. The theory appears not to be local in $x$. Instead it
depends sensitively on the {\it infrared} length scale $R$. The only
cure for this state of affairs is if there are other modes which must be
included in the effective theory. We see that this is allowed, because
the  $|m_0; p_y, p_z>$ are necessarily $x$-translation invariant 
 since they are created from the vacuum by ${\cal P}$,
which is invariant by the cyclic invariance of traces. There may be
other gapless (relative to $m_0$) states as $R \rightarrow \infty$, which
carry non-zero $x$-momentum, and therefore are not created by ${\cal
  P}$, but are nevertheless part of the full theory. They would
presumably be created by Polyakov loops with ``wiggles'', to break
$x$-translation invariance. 

A naive guess is that the point-particle
state of the effective $(2+1)$-dimensional theory remains a
point-particle state in $3+1$ dimensions. However, this does not work;
as $R \rightarrow \infty$ relativistic invariance gives the effective
theory for such a state the standard form, 
\begin{equation}
S = m \int dt  \sqrt{1 + (\partial_t x)^2 + (\partial_t
y)^2 + (\partial_t z)^2}, 
\end{equation}
which cannot match with eq. (28) without reintroducing large sensitivity to
$R$. Contrast this with the string solution,
\begin{equation}
S = \Lambda_{QCD}^2 \int dt \int_0^{2\pi R} dx~ \sqrt{1 + (\partial_t
y)^2 + (\partial_x y)^2 + (\partial_t z)^2 + (\partial_x z)^2}.  
\end{equation}
This description of a string, $y(t,x), z(t,x)$, is insensitive to large $R$,
and satisfies $(3+1)$-dimensional Poincare invariance (see section 5). 

\section{The Effective String Emerges}
 
The main plausible physical assumption in this section and the next is that 
 {\it local} probes
of the $|m_0>$ states are insensitive to the radius of the universe,
$R$.  
This is closely analogous to the fact that accelerator experiments
located on earth (local probes of the vacuum state) are insensitive to
the size of our universe. Our first goal  is to demonstrate the presence of NG
 states with non-zero
eigenvalues of $P_x$,  nearly degenerate with the $|m_0; p_y,
p_z>$.

The symmetries we will use are the spatial translations, with the 
conserved momentum generators,
\begin{equation}
P_i(t) = \int_{0}^{2 \pi R} dx \int dy \int dz~ T_{t i}(t, \vec{x}),
\end{equation}
where $T_{\mu \nu}$ is the energy-momentum tensor for the Yang-Mills theory.
Let us begin with a $|m_0; p_y,p_z>$ state, with $p_y \neq 0$ say,
\begin{equation}
P_y(t)|m_0; p_y, p_z> ~~=~~ p_y |m_0; p_y, p_z> ~~\neq~~ 0.
\end{equation}
By continuity, for small enough $x$-momentum, $q_x$,
 we have,   
\begin{equation}
\int_{0}^{2 \pi R} dx \int dy \int dz~ e^{iq_xx}~T_{t y}(t, \vec{x})
|m_0; p_y, p_z> ~~\neq~~ 0.
\end{equation}
 The operator on the left-hand side has injected
momentum $q_x$ into the $|m_0>$ state, so the result has
non-zero $x$-momentum and therefore has no overlap with any of the 
$|m_0>$ states, which were shown earlier to be $x$-translation
invariant. On the other hand for small $q_x$, by
continuity again, eq. (33) must have
a non-vanishing overlap with some energy-eigenstate whose energy
is near $\sqrt{m_0 + p_y^2 + p_z^2}$. These gapless (as $R
\rightarrow \infty$) states with non-zero $x$-momentum are the NG modes
we are seeking. 

The NG states carrying $x$-momentum cannot be obtained by Lorentz
transformations (for example an $x$-boost) from the $|m_0; p_y, p_z>$
states. The
reason is that these modes are local in $x$, and by the assumption at
the beginning of this section they should be insensitive to large $R$,
so that they are constrained by $(3+1)$-dimensional Poincare
invariance. Therefore if these states and the $|m_0; p_y, p_z>$ form a single
multiplet they must have a common mass, $m$, which is insensitive to
large $R$. This contradicts the strong $R$-dependence in $m_0$. This
argument just rephrases the case of eq. (29) disposed of in the previous 
section. We conclude that applying the operator
$\int_{0}^{2 \pi R} dx \int dy \int dz~ e^{iq_xx}~T_{t y}(t, \vec{x})$
to the $|m_0; p_y p_z>$ creates new Poincare multiplets of gapless
modes. 
Similarly, we can replace ``$y$'' by ``$z$''
in the above manipulations to get NG modes related to $z$-translations.

It is very important  to understand why we cannot inject
momenta $q_y, q_z$ in creating these NG modes. Consider the state
\begin{equation}
\int_{0}^{2 \pi R} dx \int dy \int dz~ e^{iq_yy}~T_{t y}(t, \vec{x})
|m_0; p_y, p_z> ~~\neq~~ 0, 
\end{equation}
which also reduces to eq. (32), as $q_y \rightarrow 0$. There is no
reason why this state should not be a superposition of $|m_0; p_y + q_y,
p_z>$ plus other energy eigenstates lying above some finite gap. That
is, there is no Goldstone theorem for new gapless modes carrying $q_y,
q_z$ to be  produced. 

The properties we have found are the qualitative hallmarks of the
spectrum of a
(single) string wrapped around the $x$-axis: energy eigenstates are
labelled by the transverse momenta of the whole string, $p_y, p_z$,
while further NG quanta carrying {\it only} $x$-momentum can be added,
in correspondence to the $y$ and $z$ vibrational degrees of freedom
along the string. The precise form of the string spectrum will be
deduced in the next section.

The derivation presented here is similar to Goldstone's
theorem for internal
symmetries, with the broken internal symmetry charge and current replaced by
the momentum
generators, $P_y, P_z$, and $T_{\mu \nu}$ repectively, and 
 with the non-invariant degenerate vacuum states replaced
by $|m_0, p_y, p_z>$. The main difference is that in the case of internal
symmetry breaking the ground state is invariant under
 all spacetime symmetries, so
that there are  gapless NG modes with all components of momentum,
corresponding to massless point-particle excitations. 

\section{The String Chiral Lagrangian}

It is more convenient in constructing the effective lagrangian to
translate our findings
entirely into position space. To this end, note that the 
$|m_0; p_y, p_z>$ states can be written, 
\begin{equation}
|m_0; p_y, p_z> ~~=~~ \int \int  dy dz~ e^{ip_yy + ip_zz} |m_0; y, z>,
\end{equation}
where the $|m_0; y, z>$ are the conjugate position
eigenstates for $y$ and $z$, but remain $x$-translation invariant.
Note that they are not energy eigenstates, though by taking $R$, and
hence $m_0$, to be large we can make arbitrarily long-lived and 
localized (in $y$ and
$z$) gaussian wavepackets of these states.\footnote{This is just the
  approach to the ``static limit'' $m_0 \rightarrow \infty$ in $2+1$
  dimensions, so-called because in this limit position eigenstates
  are stationary states.} 
 The derivation of the last section can therefore be repeated by
replacing $|m_0; p_y, p_z>$ by such $y,z$-localized wavepackets. Since
$T_{ty}$ is the charge density for local $y$-translations (and similarly for
$z$), we see that acting on the wavepacket localized at $y_0$ (for all $x$)
with $\int_{0}^{2 \pi R} dx \int dy \int 
dz~ e^{iq_xx}~T_{t y}(t, \vec{x})$
creates a state with $y$ localized in an $x$-dependent fashion. These
are the NG states of the last section.
 We can therefore denote the gapless NG states by fields
$y(t,x)$, $z(t,x)$, to describe the $x$-dependent
localization in $y$ and $z$, at any time $t$. This is quite analogous to
the interpretation of NG fields for internal symmetry breaking.

Thus the NG fields, $y(t,x), z(t,x)$, define a ``world-sheet'' surface
in spacetime. This simple physical interpretation allows us to
straightforwardly determine their transformations under
$(3+1)$-dimensional Poincare 
symmetry.\footnote{Of course the finite $x$ radius
explicitly breaks this symmetry, but this is not significant 
  for the local couplings of the  NG fields,  for large $R$.
 That is, $1/R$ is a ``soft'' explicit 
breaking of Poincare
symmetry. Again this is completely analogous to the fact that Lorentz
symmetry holds good in terrestrial experiments despite perhaps being
invalid for our universe globally.}
 Under the Poincare {\it sub}group associated with  the $x$ and
$t$ directions, they transform as scalar fields. Under $y$ and $z$
translations they shift, just as one would expect of NG modes associated
with the spontaneous breaking of these symmetries. Under general
Lorentz transformations $\Lambda_{\mu}^{\nu}$, the vector $(t, x, y(t,x),
z(t,x))$ transforms to $(t', x', y'(t',x'), z'(t', x'))$ by
multiplication by the tensor $\Lambda$. These results can also be
derived more laboriously from the construction of the NG modes in terms
of  the energy-momentum tensor $T_{\mu\nu}$ in the last section,
 and using the $T_{\mu\nu}$ current algebra. What is noteworthy is that,
 though the Poincare algebra implies that the non-linear realization of
 the $y, z$ translations necessarily results in a non-linear realization
 of some of the
 Lorentz transformations, we only have NG modes corresponding to $P_y,
 P_z$. The reason is that the Poincare algebra has fewer independent 
currents than it has generators: $T_{\mu\nu}$ is used to construct both
the translation and Lorentz generators.

The spacetime symmetry gives the central constraint on the construction
of the effective lagrangian. The constraint is that the action must be
constructed out of geometric invariants of the embedding of the
world-sheet surface in spacetime. The simplest and most important of
these is the surface area, 
\begin{equation}
S = \Lambda_{QCD}^2 \int dt \int_0^{2\pi R} dx~ \sqrt{1 + (\partial_t
y)^2 + (\partial_x y)^2 + (\partial_t z)^2 + (\partial_x z)^2}, 
\end{equation} 
which is just the Nambu-Goto string action. Other invariants involve
more derivatives and are consequently less important at low energies in
quantizing about the classical string ground state where $y, z$ are
constants. Now that we have derived the existence of the effective
string theory within QCD, the considerations of ref. \cite{effst} apply
to its quantization.

Expanding the Nambu-Goto term in derivatives and continuing to Minkowski
space gives,
\begin{equation}
S =  \Lambda_{QCD}^2 \int dt \int_0^{2\pi R} dx~ \{-1 + \frac{1}{2}[(\partial_t
y)^2 - (\partial_x y)^2 + (\partial_t z)^2 - (\partial_x z)^2] + ...~\}, 
\end{equation}
which makes quantization straightforward (free $(1+1)$-dimensional
scalar field theory). In the sector with
$y$ and $z$ independent of $x$ we recover the spectrum of the $|m_0>$
states found in section 3. Thus the string action eq. (36) summarizes
all the information
we have learned in previous sections 
about the lowest-lying winding glue states, plus it enforces the
constraints of Poincare invariance. 

 In the  formalism of Volkov for
non-linear realizations of spacetime symmetries \cite{volkov}
\cite{ogiev} applied to the breaking of $y,z$-translations, $y(x,t)$ and
$z(x.t)$ provide the minimal NG field content.  
In ref. \cite{hughes}, a similar construction to that in this section was
used to describe effective superstrings that emerge in the
supersymmetric abelian Higgs model.

\section{Discussion}

In this paper I have paralleled the well-known route to deriving the
existence and behavior of Nambu-Goldstone (NG) modes associated to
internal symmetries:
dynamics $\rightarrow$ (almost) degenerate (winding) 
ground states which are not
invariant under the symmetry $\rightarrow$ ``Goldstone theorem'' proving
the existence of gapless NG modes $\rightarrow$ chiral lagrangian
derivative expansion for the NG modes, constrained by the non-linear
realization of the full dynamical symmetry. The NG modes are strings,
not particles, because the ``broken'' symmetries are spacetime
symmetries, not internal symmetries.

Since this derivation of effective strings relied on $R \gg
\Lambda_{QCD}^{-1}$, it is valid as a theory of long strings,
Furthermore, since the general terms in the string effective lagrangian
are geometrical invariants of the world-sheet embedding in spacetime,
the Nambu-Goto term (surface area) only dominates for world-sheets which
are locally flat on $\Lambda_{QCD}^{-1}$ length scales. Thus there would
appear to be  a clear domain of validity for the effective theory. There
is however one possible worry. An effective theory based only on the
fact of spontaneous symmetry breaking, includes the minimal NG fields,
but does not inform us about the presence or absence of other massless
modes.\footnote{For example, in QCD, spontaneous chiral symmetry
  breaking implies
a (nearly) massless pion, but it does not tell us if the $\rho$ meson is
massless or massive enough to be omitted from the chiral lagrangian.} 
For the case of hadronic strings I have only included the
minimal NG fields in the chiral lagrangian. The more general case when
there are other gapless modes can be treated by the  formalism
of Volkov \cite{volkov} \cite{ogiev} \cite{hughes}. 
I have not yet resolved whether these extra modes will always
decouple from the string modes at low energies, analogously to the case
of internal symmetry breaking.

The remaining problem is that in this paper I only treated pure
Yang-Mills theory, describing a closed string winding around the compact
$x$-direction, whereas phenomenologically we want to consider mesons and
baryons. The important observation is that the NG string modes are {\it
  local} excitations of the stretched glue, and therefore should be
insensitive to boundary conditions if they are far-removed. Thus if a
meson is stretched by its angular momentum, the glue in between the
quarks should still be described by effective strings. The region near
the  quark ends however requires special consideration. One further
complication when dynamical quarks are introduced is that strictly there
is no confinement but rather screening,  due to quark pair
production. This can be avoided by using the large-$N_c$ limit to
suppress the pair production. Finite-$N_c$ corrections will then
correspond to the splitting of the effective strings. In a second paper,
I will treat mesons in this manner, taking care to remain in the domain
of validity of the effective string theory. Baryons are necessarily
trickier, though there are some indications that their string picture
may simplify in the large-$N_c$ limit \cite{witten}. 
Their treatment is left for the future.

\section*{Acknowledgments}
This research was supported by the U.S. Department of Energy under grant
\#DE-FG02-94ER40818.
I am grateful for useful exchanges with Sidney Coleman, Howard Georgi,
Albion Lawrence, Markus Luty, Joseph Polchinski, and Martin Schmaltz.

\end{document}